\documentclass[conference]{IEEEtran}
\IEEEoverridecommandlockouts

\usepackage{cite}
\usepackage{amsmath,amssymb,amsfonts}
\usepackage{algorithmic}
\usepackage{graphicx}
\usepackage{textcomp}
\usepackage{xcolor}
\usepackage{booktabs}
\usepackage{makecell}
\usepackage{multirow}
\usepackage{array}
\usepackage{tabularx}
\usepackage{amsmath}
\usepackage{siunitx}

\def\BibTeX{{\rm B\kern-.05em{\sc i\kern-.025em b}\kern-.08em
    T\kern-.1667em\lower.7ex\hbox{E}\kern-.125emX}}
\begin{document}

\title{How Meta-Learning Shapes LoRA Adapter Geometry in Speech Deepfake Detection}



\author{
\IEEEauthorblockN{Ivan Kukanov$^{1}$, Janne Laakkonen$^{2}$, Ville Hautam{\"a}ki$^{2}$}
\IEEEauthorblockA{
$^{1}$KLASS Engineering \& Solutions, Singapore \\
$^{2}$ University of Eastern Finland, Finland \quad}
}

\graphicspath{{imgs}}

\maketitle
\bstctlcite{IEEEexample:BSTcontrol}

\begin{abstract}

Meta-learning for domain generalization (MLDG) improves out-of-distribution speech deepfake detection over empirical risk minimization (ERM) when both objectives train low-rank adapters on the same frozen self-supervised speech model. Because the architecture and adapter capacity are held fixed, this gap points to differences in how the training objective shapes the adapter, yet the field characterizes objectives through error rates rather than through the geometry of the solution they reach. We introduce a descriptive diagnostic for this question: holding architecture, rank, data, and seeds fixed and varying only the objective, we use the empirical Fisher on the finished adapter to compare the geometry that ERM and MLDG leave behind. We characterize each adapter with effective-rank diagnostics that separate where the adapter changes from where those changes matter to the loss, resolved by projection and by depth. Applied to ERM and MLDG, the diagnostic shows that the objective does not reshape all adapter projections alike: the loss-relevant update concentrates in the query and key projections while becoming more distributed in the output projection, consistently across six corpora and most strongly in the upper layers. The same contrast appears in the merged update independently of the low-rank factorization, indicating that it reflects the geometry of the effective update rather than the parameterization. These results show that the gap between ERM and MLDG is not only a difference in error rate, but a difference in how loss-relevant capacity is organized inside the adapter, and that loss-aware adapter geometry is a way to see it.

\end{abstract}


\section{Introduction}

Speech deepfake detection is fundamentally a generalization problem~\cite{laakkonen2026generalizable}. A detector is trained on the synthesizers available today, but it will be used against the synthesizers that exist tomorrow, built from new vocoders, new text-to-speech models, or existing ones retrained from different seeds. Each generator leaves its own acoustic artifacts, so a detector that learns the artifacts of known attacks captures cues the next generator does not share, and its accuracy drops on attacks it has not seen \cite{muller22_interspeech}. Bonafide speech also varies across corpora, languages, and recording conditions, but the attack side adds a further open-ended source of shift, since each new generator can introduce artifacts unlike any seen before. This makes detection a domain-generalization problem \cite{zhou2021dgsurvey}: the detector must use cues that separate synthetic from real across generators, not the signature of any single one.

\begin{figure}[htbp]
\centering
\includegraphics[scale=0.55, trim={0cm 0.9cm 0 0.9cm}, clip]{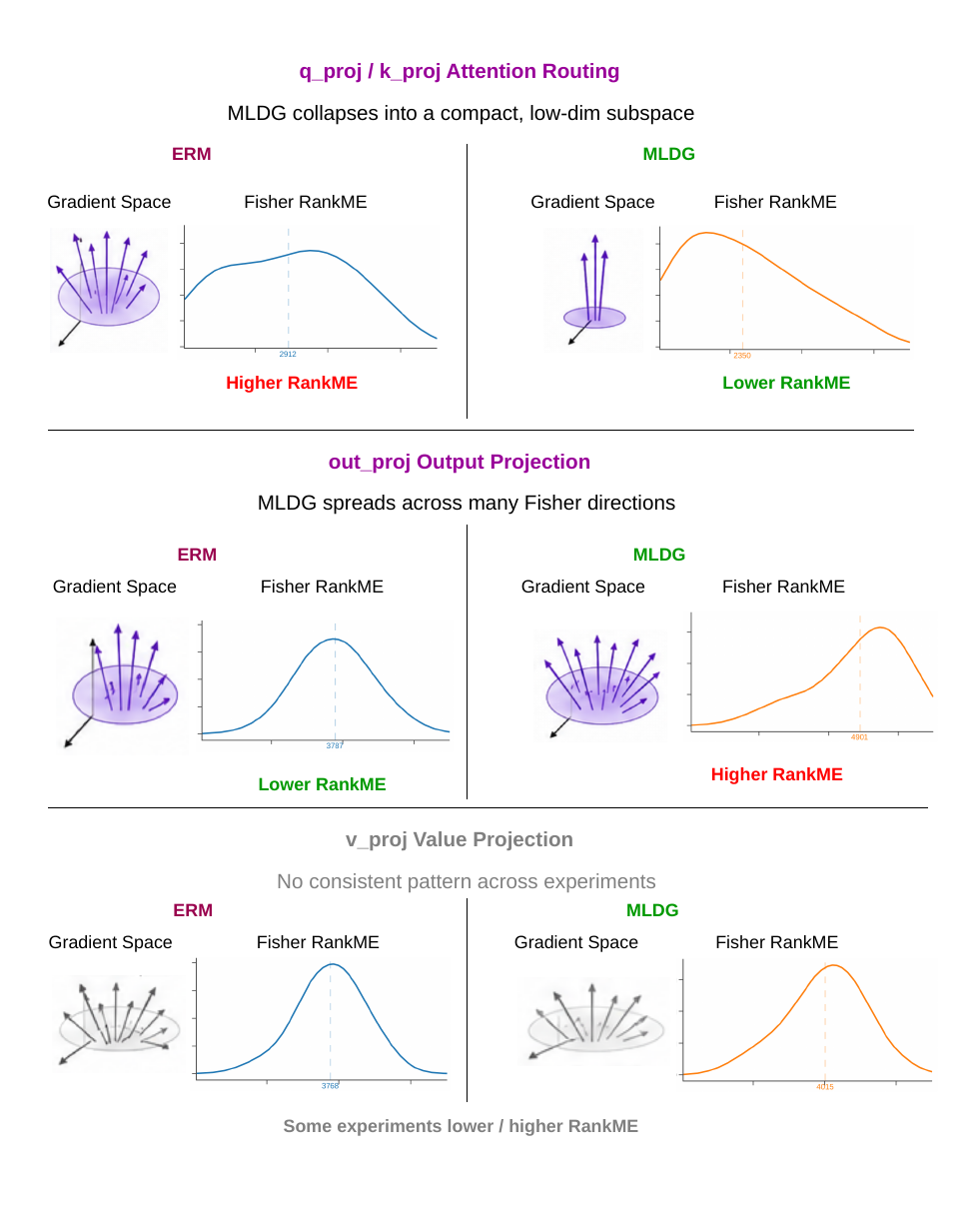}
\caption{ERM- vs.\ MLDG-trained LoRA adapters in a frozen {Wav2Vec-AASIST} deepfake detector, compared via Fisher effective rank $\mathrm{RankME}_{F}$ diagnostic: MLDG reallocates Fisher energy, gradient space of query/key projections are concentrated, while output projection is distributed (purple color). The effect is consistent across \emph{seeds} $\times$ \emph{layers} $\times$ \emph{corpora}. The effect is faded for value projection (gray color)}
\label{fig:rankme_heatmap}
\end{figure}

One way to address this problem is meta-learning \cite{thrun1998learning}. A common use of meta-learning is few-shot learning or adaptation: rather than training a model to solve one task, it trains across many tasks to produce a model that can adapt to a new task quickly, using only a few examples and a few gradient steps. In gradient-based meta-learning, this is often implemented as a bilevel loop: an inner step adapts the model to a task, and an outer objective rewards parameters from which that adaptation succeeds \cite{finn2017model, Kukanov_2024}. Meta-learning for domain generalization (MLDG)~\cite{Li2018MLDG, laakkonen2026generalizable} re-purposes this machinery for domain shift rather than few-shot adaptation. It treats domains as the tasks: at each step it adapts on a subset of training domains and then evaluates that adapted state on held-out domains, so the objective favors updates that transfer to domains the model was not adapted on \cite{Li2018MLDG}. Recent work applies MLDG to speech deepfake detection in a parameter-efficient setting, training only small low-rank adapters (LoRA) \cite{hu2022lora} inserted into a frozen self-supervised speech model \cite{laakkonen2026generalizable}. At a fixed adapter budget, replacing standard training, or empirical risk minimization (ERM) \cite{Vapnik1991PrinciplesOR}, with MLDG substantially improves out-of-distribution detection without adding capacity, pointing to the training objective, rather than adapter size, as the source of the gain \cite{laakkonen2026generalizable}.

The prior study reported this gain through error rates \cite{laakkonen2026generalizable}, without examining the geometry of the solution each objective produces, and left the analysis of what the adapters learn as an open direction. The question is therefore not whether the objective matters, which is established, but how it reshapes the adapter: \emph{given the same architecture, adapter capacity, and data, how does training with MLDG rather than ERM change the geometry of the learned adapter?}

Answering this needs a way to read a finished adapter, not a way to train a better one. We treat the trained adapter as a displacement from its initialization and ask, for each attention projection and each layer, how concentrated that displacement is, both on its own and after weighting each direction by how much the loss depends on it. This separates where the adapter changed from where those changes matter. Holding architecture, rank, data, and seeds fixed and varying only the objective, the comparison isolates the effect of the objective alone. The diagnostic combines two established tools, the empirical Fisher \cite{martens2020new} and effective rank \cite{roy2007effective,garrido2023rankme}, but uses them descriptively rather than to define or guide the training objective.

This paper makes two contributions:
\begin{itemize}
\item We introduce a descriptive, objective-comparative diagnostic of adapter geometry: holding the parameterization fixed and varying only the objective, we use the empirical Fisher and effective rank to localize, by projection and by depth, where a training objective places the loss-relevant part of the adapter update.

\item Applying this diagnostic to ERM and MLDG, we find that MLDG concentrates the loss-relevant update in the query and key projections while distributing it in the output projection, rather than reshaping the adapter uniformly. This pattern also appears after merging the LoRA factors into the effective update $\Delta W$, reducing the concern that it is only an artifact of the factor coordinates.
\end{itemize}

Together, these results show that the out-of-distribution gap between ERM and MLDG is not only a difference in error rate, but a difference in how the loss-relevant capacity of the adapter is organized. Weighting the adapter update by loss sensitivity makes this loss-relevant organization explicit, whereas raw update magnitude alone does not.

\section{Related Work}

\textbf{Low-rank adaptation and the geometry of trained adapters.}
LoRA~\cite{hu2022lora} freezes the pretrained model and learns an additive low-rank update, motivated by evidence that task adaptation often lies in a low-dimensional subspace~\cite{aghajanyan2021intrinsic}. Much of the LoRA literature studies how to allocate, initialize, or regularize this capacity~\cite{zhang2023adaptive,meng2024pissa}.
A separate line treats the trained adapter itself as the object of study, examining its singular structure, factor asymmetry, or relation to forgetting~\cite{shuttleworth2026lora,zhu2024asymmetry,biderman2024lora}. These analyses read a finished adapter under a single training objective; they do not ask how the same adapter parameterization changes when only the objective is varied.

\textbf{Fisher and effective-rank diagnostics.}
Recent adapter methods use Fisher, curvature, or related gradient information
constructively to precondition LoRA updates, select initialization directions, or steer rank allocation~\cite{zhang2024riemannian,feng2026learning,zheng2026curvature}, all of which shape adaptation before or during training. We instead use the empirical Fisher descriptively, on the finished adapter, reading it as a loss-sensitivity summary of the trained adapter rather than as a curvature estimate for optimization~\cite{kunstner2019limitations}. To quantify this concentration, we use entropy-based effective rank, which measures how concentrated a distribution or spectrum is~\cite{roy2007effective} and has been applied to representation quality as RankMe~\cite{garrido2023rankme}. Our starting point is MLDG-LoRA for speech deepfake detection~\cite{laakkonen2026generalizable}, where replacing empirical risk minimization with a meta-learning objective improved cross-condition generalization without increasing adapter capacity, but the learned adapter itself was not analyzed. To our knowledge, prior work has not used descriptive, objective-comparative geometry to study how a training objective reshapes a fixed low-rank adapter in a frozen self-supervised speech model.

\section{Methodology}


%

We analyze the geometry of low-rank adaptation under empirical risk minimization (ERM) and meta-learning for domain generalization (MLDG). Using the detector architecture of~\cite{laakkonen2026generalizable} as a fixed testbed, we ask a question: \emph{How does the training objective shape the geometry of the learned LoRA adapters?} The same rank-$r$ ($r = 16$) LoRA parameterization is inserted into a frozen Wav2Vec 2.0--AASIST backbone~\cite{baevski2020wav2vec,jung2022aasist} and trained twice: once with ERM and once with first-order MLDG~\cite{Li2018MLDG}.

Let $f_{\theta_0,\phi}$ denote a frozen Wav2Vec backbone with LoRA parameters $\phi$ inserted into the self-attention projections (q, k, v, and out). For each adapted linear block $W_\ell \in \mathbb{R}^{d_{\text{out}}\times d_{\text{in}}}$, LoRA parameterizes the update as
\begin{equation}
W_\ell' = W_\ell + \Delta W_\ell, \qquad \Delta W_\ell = B_\ell A_\ell,
\end{equation}
where $A_\ell \in \mathbb{R}^{r\times d_{\text{in}}}$ and $B_\ell \in \mathbb{R}^{d_{\text{out}}\times r}$ are trainable low-rank factors and $r \ll \min(d_{\text{in}}, d_{\text{out}})$ \cite{hu2022lora}. We analyze the learned adapter displacement
\begin{equation} \label{eq:displacement}
\Delta \phi = \phi - \phi_0,
\end{equation}
where $\phi_0 = [\operatorname{vec}(A_{\text{init}}), \operatorname{vec}(B_{\text{init}})]$ vectorized and concatenated LoRA initializations, $A_{\text{init}}$ is drawn from a scaled Gaussian and $B_{\text{init}}$ is initialized to zero, and $\phi = [\operatorname{vec}(A), \operatorname{vec}(B)]$ - trained LoRA weights.

\subsection{Training Objectives}

Given source domains $\mathcal{S} = \{\mathcal{D}_1, \dots, \mathcal{D}_M\}$ and task loss $\ell(\cdot)$, pooled-source ERM minimizes
\begin{equation}
\mathcal{L}_{\mathrm{ERM}}(\phi) = \sum_{d \in \mathcal{S}} \mathbb{E}_{(x,y)\sim\mathcal{D}_d}\!\left[\ell\big(f_{\theta_0,\phi}(x), y\big)\right].
\end{equation}
MLDG simulates train--test domain shift inside each iteration by splitting source domains into meta-train and meta-test subsets and optimizing
\begin{equation}
\mathcal{L}_{\mathrm{MLDG}}(\phi)
= \mathcal{L}_{\mathrm{tr}}(\phi)
+ \beta \, \mathcal{L}_{\mathrm{te}}\!\left(\phi - \alpha \nabla_\phi \mathcal{L}_{\mathrm{tr}}(\phi)\right),
\end{equation}
where $\alpha$ and $\beta$ denote the inner-step size and meta-test weighting coefficient, respectively \cite{Li2018MLDG,zhou2021dgsurvey}.

\subsection{Diagnostic: Fisher-Weighted Geometry of LoRA Updates}

To quantify where the loss-relevant adapter update is concentrated, we weight the learned displacement $\Delta \phi$ from (\ref{eq:displacement}) by an empirical Fisher estimate over the LoRA parameters only. For a probe set $\mathcal{D}_{\mathrm{probe}}=\{(x_i,y_i)\}_{i=1}^{N}$, we accumulate squared per-example gradients to obtain the diagonal of the empirical Fisher,
\begin{equation} \label{eq:fisher_diag}
\widehat{F}_p
=
\frac{1}{N}
\sum_{i=1}^{N}
\left(
\frac{\partial \ell\big(f_{\theta_0,\phi}(x_i), y_i\big)}
{\partial \phi_p}
\right)^2,
\end{equation}
for each LoRA parameter $\phi_p$, $p=1,\dots,P$. We use the diagonal estimate
$\widehat{F}\in\mathbb{R}^{P}$ rather than the full $P\times P$ Fisher; it is a practical curvature diagnostic motivated by natural-gradient geometry, and we interpret it explicitly as an empirical-Fisher approximation rather than the exact Fisher metric \cite{Martens2014NaturalGradient,Martens2015KFAC,Wu2024ImprovedEmpiricalFisher}.

We define the Fisher-weighted importance of each LoRA coordinate as
\begin{equation}
\pi_p
=
\frac{\widehat{F}_p\,\Delta\phi_p^{2}}
{\sum_{q=1}^{P}\widehat{F}_q\,\Delta\phi_q^{2}},
\end{equation}
which is large where the adapter both moved and the loss is sensitive. The importance is computed per adapted block, with $p$ ranging over the concatenated $A_\ell$ and $B_\ell$ factors of that block.

We summarize the importance distribution $\{\pi_p\}_{p=1}^{P}$ with three complementary statistics:
\begin{align}
C_k &= \sum_{p=1}^{k}\pi_{(p)}, \\
E(m) &= \sum_{p=1}^{m}\pi_{(p)}, \\
\mathrm{RankME}_{F}
&=
\exp\!\left(
-\sum_{p=1}^{P}\pi_p\log(\pi_p+\varepsilon)
\right),
\end{align}
where $\pi_{(p)}$ denotes the importance sorted in descending order, $C_k$ measures concentration in the top-$k$ coordinates, $E(m)$ is the cumulative importance curve, and $\mathrm{RankME}_{F}$ is the effective rank of the importance spectrum \cite{Ghorbani2019HessianDensity,Jacot2018NTK,Garrido2022RankMe, BenArous2023OutlierEigenspaces}. A lower $\mathrm{RankME}_{F}$ indicates that the loss-relevant update is carried by fewer coordinates.

\begin{table*}[!htbp]
\caption{Per-module $\mathrm{RankME}_F$ of the LoRA trained under ERM vs.\ MLDG, averaged over 5 seeds, 6 corpora, and 24 layers
(higher - energy spread over more Fisher directions). $\Delta=\mathrm{MLDG}-\mathrm{ERM}$;
\emph{Seeds agree} counts sign-consistent seeds; \emph{Consistency} is the descriptive seed$\times$dataset count. MLDG concentrates $W_q$/$W_k$ (${\approx}-19\%$) and distributes $W_{\mathrm{out}}$ (${+}29\%$), both $5/5$ seeds.}
\begin{center}
\begin{tabular}{c c c c c c c c}
\hline
\textbf{Module} & \textbf{ERM $\mathrm{RankME}_F$} & \textbf{MLDG $\mathrm{RankME}_F$} & \textbf{$\Delta$} & \textbf{$\Delta\%$} & \textbf{Effect} & \textbf{Seeds agree} & \textbf{Consistency} \\
\hline
\hline
q\_proj   & 3174 & 2559 & $-615$  & $-19.4\%$ & concentrate $\downarrow$ & 5/5 & 30/30 \\
k\_proj   & 2912 & 2350 & $-562$  & $-19.3\%$ & concentrate $\downarrow$ & 5/5 & 30/30 \\
v\_proj   & 3768 & 4015 & $+247$  & $+6.6\%$  & distribute $\uparrow$    & 3/5 & 18/30 \\
out\_proj & 3787 & 4901 & $+1114$ & $+29.4\%$ & distribute $\uparrow$    & 5/5 & 30/30 \\
\hline
\end{tabular}
\label{tab:erm_mldg_module_effect}
\end{center}
\end{table*}

\subsection{Evaluation protocol} \label{sec:eval_proto}

We evaluate these diagnostics under cross-domain audio deepfake detection. Following prior work~\cite{laakkonen2026generalizable}, each dataset or generator family is treated as a domain, and we perform leave-one-domain-out evaluation over benchmarks such as ASVspoof~2019 LA Eval~\cite{Todisco2019ASVspoof2F}, ASVspoof~2021 LA and DF \cite{Yamagishi2021ASVspoof2021}, ASVspoof~5~\cite{Wang2024ASVspoof5C}, InTheWild \cite{muller22_interspeech}, FakeAVCeleb \cite{Khalid2021FakeAVCeleb}. We report Fisher alignment, domain-wise Fisher stability, and output-space effective rank per layer and projection type, and then correlate these diagnostics with out-of-domain detection metrics. This protocol is motivated by the well-documented generalization gap of audio deepfake detectors across datasets, attacks, and recording conditions \cite{Yi2023AudioDeepfakeSurvey,muller22_interspeech,Hsu2021RobustWav2Vec2}.

\section{Experimental Setup}

\subsection{Backbone and adapters} All diagnostics are computed on the checkpoints from~\cite{laakkonen2026generalizable}: frozen self-supervised (SSL) speech encoder Wav2Vec~2.0 XLSR-53 front-end (24 transformer layers, $1024$-dim embeddings); followed by AASIST back-end spectro-temporal graph-attention classifier with a binary (bonafide/spoof) output trained under a negative log-likelihood objective. The SSL encoder and the classifier head are kept frozen; the probing diagnostic was done on pretrained LoRA adapters. LoRA adapters of rank $r{=}16$ and scaling $\alpha_{\mathrm{LoRA}}{=}2$ inserted into the query, key, value, and output projections of every self-attention block of SSL ($4 \times 24 = 96$ adapters). Training objectives are pooled-source ERM or first-order MLDG~\cite{Li2018MLDG}. We analyze the five released seeds $\{42, 123, 555, 999, 2023\}$, yielding five paired ERM/MLDG checkpoints~\cite{laakkonen2026generalizable}. For a fixed seed, the LoRA adapter displacement $\Delta\phi = \phi - \phi_0$ equation (\ref{eq:displacement}) is frozen.

\subsection{Probe Dataset for Empirical Fisher and Alignment}

For each trained checkpoint, we characterize how the learned LoRA update $\Delta\phi$ aligns with the loss curvature through the empirical Fisher information equation (\ref{eq:fisher_diag}). We adopt the diagonal Fisher $\widehat{F}_p$, estimated by accumulating squared gradients of the LoRA parameters only, the SSL backbone and AASIST are frozen. 

Restricting attention to the diagonal amounts to measuring alignment in the coordinate basis ($\Delta\phi_p$, $\widehat{F}_p$) and induces a per-parameter importance $\pi_p \propto \widehat{F}_p\,\Delta\phi_p^2$, normalized to a distribution over the adapter coordinates $\pi_p$. The update $\Delta\phi_0$ is taken per-seed checkpoint initialization. All alignment statistics are computed per \textit{seed} $\times$ \textit{dataset} $\times$ \textit{projection} $\times$ \textit{layer} in Fig. \ref{fig:rankme_heatmap}.

Adapters from~\cite{laakkonen2026generalizable} are trained on ASVspoof 2019 LA, and the Fisher information is estimated on six evaluation corpora from Sec. \ref{sec:eval_proto}, spanning a range of distribution shifts.
Because the importance $\pi_p$ depends only on the frozen update and the curvature, each corpus is subsampled to a class-balanced set (1000 bonafide and 1000 spoof utterances) to remove class-prior effects from the estimate; inputs are raw 16KHz waveforms cropped to $\approx 4$s windows.

From the sorted importance distribution $\{\pi_p\}$ of each module we report the Fisher effective rank $\mathrm{RankME}_F$, which is high when the update spreads across many curvature directions and low when it concentrates.

\section{Results}

\subsection{Fisher Alignment: ERM vs MLDG LoRA Updates}

Analysis of Fisher $\mathrm{RankME}_F$ and concentration metrics across six evaluation datasets from Sec.~\ref{sec:eval_proto} 
and five seeds $\{42, 123, 555, 999, 2023\}$, which is $30$ runs, on the rank-$16$ LoRA adapters of a $24$-layer Wav2Vec 2.0--AASIST model~\cite{laakkonen2026generalizable}, see Tab.~\ref{tab:erm_mldg_module_effect} and Fig.~\ref{fig:rankme_heatmap}. 

Empirically, there is no significant global difference between ERM and MLDG in terms of the globally averaged $\mathrm{RankME}_F$ metric (ERM $\mathrm{RankME}_F$ = 3410, MLDG $\mathrm{RankME}_F$ = 3456). ERM and MLDG spend roughly the same total estimated Fisher energy, but MLDG relocates it across projection modules and layer depth. Effects are extremely consistent across seeds, Tab \ref{tab:erm_mldg_module_effect}. The q\_proj and k\_proj are concentrated and out\_proj is distributed, effects are consistent across \textit{seed} $\times$ \textit{dataset} (30/30 experiments). While v\_proj trend is noisier (18/30 experiments). 

\textit{Depth structure of the reallocation (per-module $\times$ depth)}. Splitting each
projection into lower~(layers 0–11) and upper~(12–23) shows the reallocation is not
flat with depth layers, Tab.~\ref{tab:rankf_per_module_depth}. The q/k \textit{concentrate} effect sharpens toward the top: q\_proj -10.9\% lower vs -33.0\% upper, k\_proj -13.5\% vs -30.8\% upper (both 5/5 seeds). The attention routing is compressed hardest in the task-specific upper layers. The out\_proj \textit{distribute} effect runs the other way, strongest low and fading up: +42.2\% lower (5/5) vs +15.8\% upper (4/5). V\_proj distributes only in the lower layers (+15.3\%, 5/5) and is flat/noisy at the top (-3.0\%, 3/5). The MLDG spreads generic acoustic content across many directions in the lower layers while tightening task-specific routing into few directions high up.

\textit{The reallocation holds in every dataset (per-dataset $\times$ per-module)}. Broken out by evaluation sets Tab. \ref{tab:rankf_per_dataset}, the per-module signs are identical across all 6 datasets - q and k negative everywhere (q -15\% to -23\%, k -16\% to -22\%), out\_proj strongly positive everywhere (+26\% to +32\%), v\_proj weakly positive everywhere (+3\% to +10\%). No dataset flips any module's direction. Since displacement $\Delta \phi$ is frozen within a seed and only the Fisher estimate changes across datasets, this per-dataset uniformity is descriptive corroboration that the q/k-concentrate + out\_proj-distribute structure is a property of the learned adapter, not of the data used to probe it.

\begin{table*}[htbp]
\caption{Per-module $\times$ per-depth $\mathrm{RankME}_F$ under ERM vs.\ MLDG, split into lower (0--11) and upper (12--23) backbone layers, averaged over 5 seeds and 6 corpora. $\Delta=\mathrm{MLDG}-\mathrm{ERM}$; \emph{Seeds agree} counts sign-consistent seeds; \emph{Consistency} is the descriptive seed$\times$dataset count. The $W_q$/$W_k$ concentration and $W_{\mathrm{out}}$ distribution both intensify with depth, and \emph{joint} statistics MLDG shifts distributed adaptation toward the lower acoustic layers while tightening the upper task-specific ones.}
\begin{center}
\begin{tabular}{c c r r r r c c c }
\hline
\textbf{Module} & \textbf{Layer range} & \textbf{ERM} & \textbf{MLDG} & \textbf{$\Delta$} & \textbf{$\Delta\%$} & \textbf{Effect} & \textbf{Seeds agree} & \textbf{Consistency} \\
\hline
\hline
\multirow{2}{*}{q\_proj}
 & lower (0--11)  & 3913 & 3487 & $-427$  & $-10.9\%$ & concentrate $\downarrow$ & 5/5 & 25/30 \\
 & upper (12--23) & 2435 & 1631 & $-803$  & $-33.0\%$ & concentrate $\downarrow$ & 5/5 & 29/30 \\
\hline
\multirow{2}{*}{k\_proj}
 & lower (0--11)  & 3866 & 3344 & $-522$  & $-13.5\%$ & concentrate $\downarrow$ & 5/5 & 27/30 \\
 & upper (12--23) & 1958 & 1356 & $-602$  & $-30.8\%$ & concentrate $\downarrow$ & 5/5 & 29/30 \\
\hline
\multirow{2}{*}{v\_proj}
 & lower (0--11)  & 3932 & 4533 & $+601$  & $+15.3\%$ & distribute $\uparrow$    & 5/5 & 28/30 \\
 & upper (12--23) & 3604 & 3497 & $-107$  & $-3.0\%$  & concentrate $\downarrow$ & 3/5 & 16/30 \\
\hline
\multirow{2}{*}{out\_proj}
 & lower (0--11)  & 3900 & 5548 & $+1648$ & $+42.2\%$ & distribute $\uparrow$    & 5/5 & 30/30 \\
 & upper (12--23) & 3674 & 4254 & $+581$  & $+15.8\%$ & distribute $\uparrow$    & 4/5 & 24/30 \\
\hline
\hline
\multirow{2}{*}{joint}
 & lower (0--11)  & 3903 & 4228 & +325 & +8.3\% & distribute $\uparrow$ & 4/5 & 20/30 \\
 & upper (12--23) & 2918 & 2685 & -233 & -8.0\% & concentrate $\downarrow$ & 4/5 & 21/30 \\
\hline
\end{tabular}
\label{tab:rankf_per_module_depth}
\end{center}
\end{table*}

\begin{table*}[htbp]
\caption{Per-dataset $\times$ per-module $\mathrm{RankME}_F$ under ERM vs.\ MLDG across the 6 evaluation corpora, averaged over 5 seeds and
24 layers. $\Delta\%=(\mathrm{MLDG}-\mathrm{ERM})/\mathrm{ERM}$. The reallocation pattern - $W_q$/$W_k$ concentrate (${\approx}-15\%$ to $-23\%$), $W_{\mathrm{out}}$
distributes (${\approx}+26\%$ to $+32\%$) - holds on every corpus, indicating a dataset-invariant signature of the MLDG objective rather than a shift-specific
artifact.}
\begin{center}
\footnotesize
\setlength{\tabcolsep}{4pt}
\begin{tabular}{c c c c c | c c c c | c c c c}
\hline
\multirow{2}{*}{\textbf{Dataset}} & \multicolumn{4}{c|}{\textbf{ERM}} & \multicolumn{4}{c|}{\textbf{MLDG}} & \multicolumn{4}{c}{\textbf{$\Delta\%$}} \\
\cline{2-13}
 & \textbf{q} & \textbf{k} & \textbf{v} & \textbf{out} & \textbf{q} & \textbf{k} & \textbf{v} & \textbf{out} & \textbf{q} & \textbf{k} & \textbf{v} & \textbf{out} \\
\hline
\hline
ASV19 LA Eval  & 3052 & 2795 & 3564 & 3612 & 2551 & 2359 & 3906 & 4702 & $-16.4\%$ & $-15.6\%$ & $+9.6\%$ & $+30.2\%$ \\
ASV21 LA    & 2993 & 2732 & 3799 & 3885 & 2394 & 2154 & 4000 & 4934 & $-20.0\%$ & $-21.2\%$ & $+5.3\%$ & $+27.0\%$ \\
ASV21 DF    & 3031 & 2794 & 3823 & 4018 & 2384 & 2181 & 3935 & 5068 & $-21.3\%$ & $-21.9\%$ & $+2.9\%$ & $+26.1\%$ \\
ASV5    & 3298 & 2910 & 3713 & 3762 & 2540 & 2287 & 3941 & 4965 & $-23.0\%$ & $-21.4\%$ & $+6.1\%$ & $+32.0\%$ \\
InTheWild & 3411 & 3189 & 4035 & 3878 & 2894 & 2687 & 4368 & 5043 & $-15.2\%$ & $-15.7\%$ & $+8.3\%$ & $+30.0\%$ \\
FakeAVCeleb         & 3260 & 3052 & 3673 & 3568 & 2592 & 2432 & 3940 & 4696 & $-20.5\%$ & $-20.3\%$ & $+7.3\%$ & $+31.6\%$ \\
\hline
\end{tabular}
\label{tab:rankf_per_dataset}
\end{center}
\end{table*}

\subsection{Decomposing the Fisher-Weighted Effect}

\begin{figure}[htbp]
  \centering
  \includegraphics[width=0.70\columnwidth]{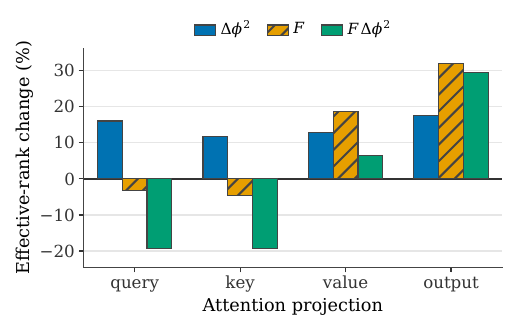}
  \caption{Per-module effective-rank change under MLDG versus ERM,
  $100\times(\mathrm{MLDG}-\mathrm{ERM})/\mathrm{ERM}$. The measures isolate the ingredients of $\mathrm{RankME}_F$: the update magnitude $\Delta\phi^2$, the empirical Fisher $F$, and their product $F\Delta\phi^2$, which is $\mathrm{RankME}_F$ itself (Table~\ref{tab:erm_mldg_module_effect}). The Fisher-based measures $F$ and $F\Delta\phi^2$ are means over 30 dataset--seed cells (six corpora, five seeds); the magnitude measure $\Delta\phi^2$ is data-free and reflects five seeds.}
  \label{fig:decomp}
\end{figure}

\begin{figure}[htbp]
  \centering
  \includegraphics[width=0.70\columnwidth]{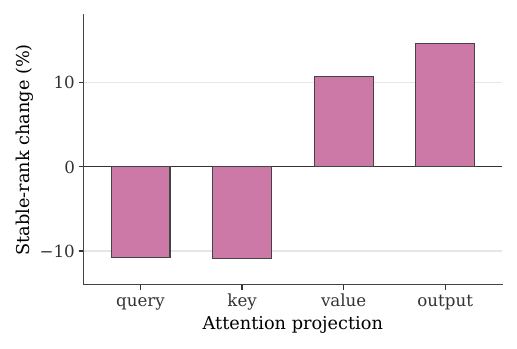}
  \caption{Stable rank of the merged update $\Delta W = \alpha AB$,
  $100\times(\mathrm{MLDG}-\mathrm{ERM})/\mathrm{ERM}$. This measure uses neither
  the Fisher nor any data and is invariant to the LoRA factor coordinates;
  values are means over five seeds.}
  \label{fig:deltaW}
\end{figure}
The per-module $\mathrm{RankME}_F$ contrast combines two factors: how much the
adapter moved and how much the loss depends on where it moved. To separate them,
we recomputed the effective rank from each factor alone: the squared update
magnitude $\Delta\phi^2$, the empirical Fisher $F$, and their product
$F\Delta\phi^2$ (the full diagnostic). Figure~\ref{fig:decomp} reports the
percent change from ERM to MLDG for each.

The magnitude-only effective rank changed in the opposite direction from the full
diagnostic on the query and key projections. Under the magnitude-only measure,
MLDG raised the effective rank of every projection ($+16.0\%$ query, $+11.6\%$
key, $+12.7\%$ value, $+17.4\%$ output), with the positive direction reproduced
across all five seeds. Under the Fisher-weighted measure, the query and key
projections fell ($-19.4\%$ and $-19.3\%$). The concentration of the
loss-relevant update on the query and key projections is therefore not a
reduction in how far those projections moved: the raw update broadened while the
Fisher-weighted update concentrated.

The output projection moved in the distributing direction under all three
measures, increasing most strongly under the Fisher-only and Fisher-weighted
measures ($+31.8\%$ and $+29.4\%$). For the Fisher-weighted measure, the query
and key signs were unanimous across the 30 dataset--seed cells. The value
projection was the only low-agreement case, with the sign matching the aggregate
direction in 18 of 30 cells.

The same query/key decrease and value/output increase also appeared in the stable
rank of the merged update $\Delta W$ (Figure~\ref{fig:deltaW}), which requires no
Fisher estimate and no probe data. MLDG lowered the query and key stable rank
($-10.8\%$ and $-10.9\%$) and raised the value and output stable rank
($+10.7\%$ and $+14.6\%$). Because this measure is invariant to the LoRA factor
coordinates, the projection-wise reorganization is present in the effective
update itself and not only in the particular $A$ and $B$ factor coordinates.

\subsection{Local Score Sensitivity to LoRA Perturbations}
We next examined whether the adapter difference is reflected in the detector's local score function. For each trained checkpoint, we added random perturbations of relative radius $\epsilon$ to the front-end LoRA factors, scaling a unit-norm random direction to $\lVert\delta\rVert_2 = \epsilon\,\lVert\phi_{\mathrm{LoRA}}\rVert_2$, held the AASIST back-end fixed, and measured the variance of the output score over five perturbation
directions on In-The-Wild. This probes local sensitivity in the trained LoRA
parameterization and is not gauge-invariant; it is not used to train or select the model.

\begin{figure}[t]
\centering
\includegraphics[width=0.80\columnwidth]{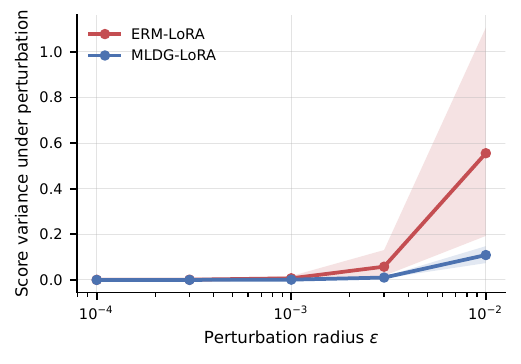}
\caption{LoRA perturbation sensitivity on In-The-Wild. Random perturbations of relative radius $\epsilon$ are applied to the trained front-end LoRA factors, with the AASIST back-end fixed, and the score variance over five perturbation directions is measured. The ERM adapter is more sensitive than the MLDG adapter at every nonzero radius. Lines show seed means; shaded regions show the seed min--max range.}
\label{fig:lora_perturbation_sensitivity}
\end{figure}

The ERM adapter was more sensitive to this perturbation than the MLDG adapter at every nonzero radius (Fig.~\ref{fig:lora_perturbation_sensitivity}). At the largest radius ($\epsilon=10^{-2}$), the mean score variance across five matched seeds was $0.56$ for ERM and $0.11$ for MLDG, a factor of $5.1$. Across the tested nonzero radii, the ERM-to-MLDG variance ratio ranged from $5.1$ to $11.0$, with ERM more sensitive throughout. The objective difference seen in the adapter geometry is therefore accompanied by a functional difference: the MLDG adapter yields a score function that is less sensitive to local LoRA perturbations on this out-of-domain corpus.

\subsection{Why MLDG Generalizes Better}

MLDG's bilevel objective (adapt on \textit{meta-train domains}, score the gradient on a held-out \textit{meta-test domain}) explicitly rewards update directions that transfer to an unseen domain and penalizes domain-specific ones. That objective leaves two complementary fingerprints in Fisher space:

\begin{itemize}
    \item Attention routing (q\_proj/k\_proj) collapses into a compact, low-dimensional subspace. MLDG commits to a parsimonious ''where to look`` pattern carried by few Fisher directions, shared across domains. A low-rank routing that survives the meta-test split is, by construction, domain-invariant. ERM has no such pressure and spreads q/k across more directions.
   \item Content written into the residual stream (out\_proj) is spread across many Fisher directions. This is the classic robustness signature: MLDG does not rely on a handful of high-curvature directions to transform representations. In anti-spoofing, those few dominant directions are where dataset-specific artifacts/shortcuts (codec, channel, silence cues) get encoded. ERM concentrates out\_proj energy there - faster training-loss reduction, but brittle off-distribution.
   \item Depth allocation: MLDG pushes distributed adaptation into the lower/mid acoustic layers (generic, transferable features) and tightens the top task-specific layers. 
\end{itemize}

ERM and MLDG use comparable total Fisher energy, but MLDG relocates it - concentrating attention routing (q/k) into an invariant low-rank subspace while diffusing content updates (out\_proj) away from the few shortcut-encoding directions. That trade of stable routing and distributed, non-shortcut content is the mechanistic reason MLDG transfers better.

\begin{figure}[htbp]
\centering
\includegraphics[scale=0.4, trim={1cm 1.4cm 0 2cm}, clip]{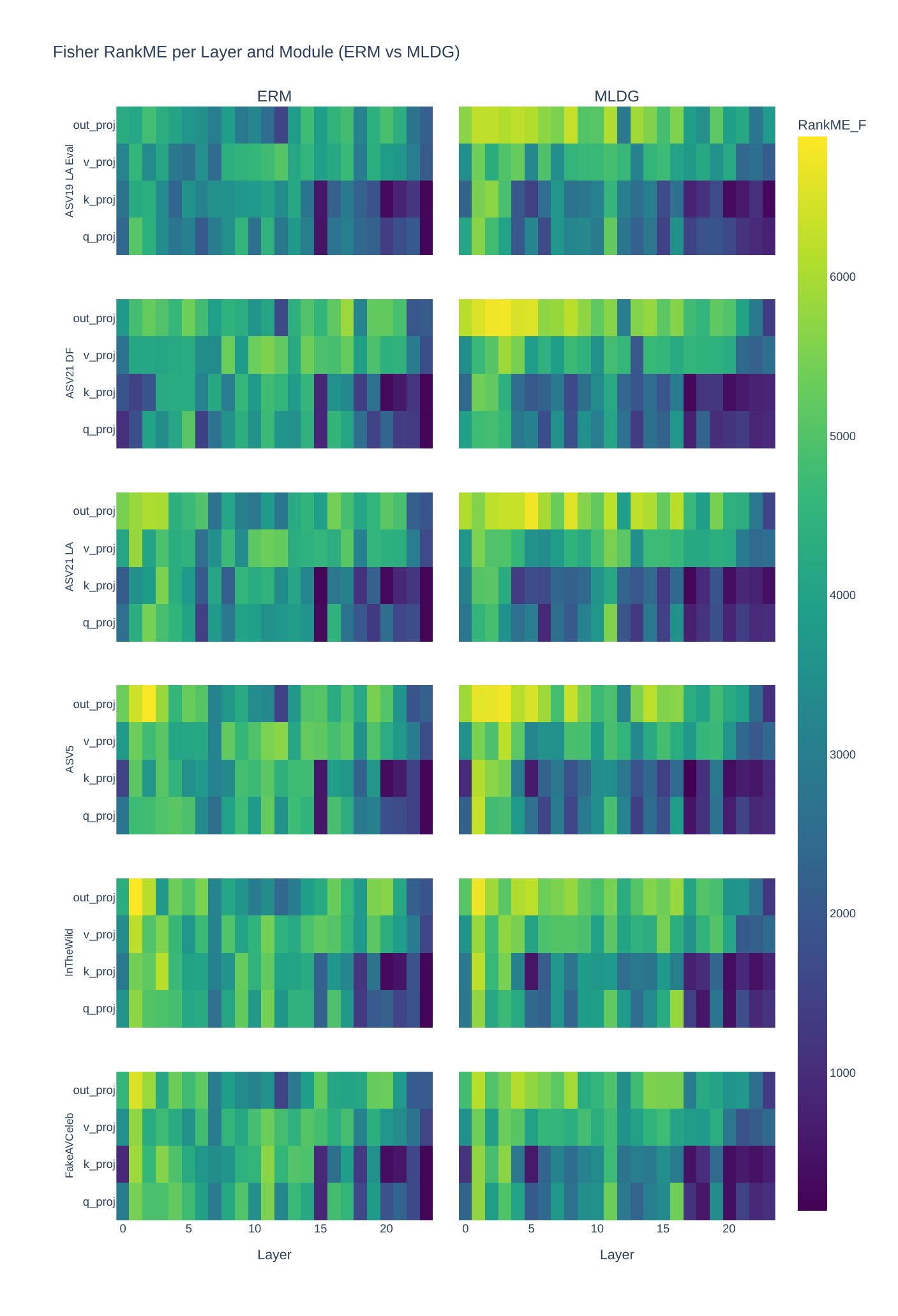}
\caption{Effect of MLDG vs ERM training, measured using RankME\_F metric with estimated Fisher.}
\label{fig:rankme_heatmap}
\end{figure}

\section{Discussion and Conclusions}
We introduced a descriptive, objective-comparative diagnostic of adapter geometry: holding architecture, rank, data, and seeds fixed and varying only the training objective, we used the empirical Fisher and effective rank to localize where an objective places the loss-relevant part of a LoRA update. Applied to ERM and MLDG in speech deepfake detection, the diagnostic showed that MLDG does not reshape the adapter uniformly. It concentrates the loss-relevant update in the query and key projections and distributes it in the output projection, consistently across six corpora, with query/key concentration strongest in the upper layers. The same contrast appeared after
merging the LoRA factors into the effective update $\Delta W$, reducing the concern that the pattern is only an artifact of the factor coordinates. The two objectives also differed in the local sensitivity of the score function they induce.

These observations suggest one interpretation of what the meta-learning objective does to the adapter. Concentrating the query and key updates into a lower-dimensional loss-relevant subspace is consistent with MLDG committing to a compact attention-routing pattern that survives the meta-test split. The distributed output update is consistent with spreading loss-relevant capacity across more directions. Whether this reorganization is what drives the out-of-distribution gain, however, is not established by geometry alone: the diagnostic is descriptive, and the perturbation result is a single functional
probe on one corpus. The reorganization is a robust property of the learned adapter, but its causal link to generalization remains a hypothesis for future work.

The diagnostic itself is not tied to this objective pair or to speech. It compares the geometry that any two training objectives leave in a fixed parameterization, and could be applied to other adaptation objectives, other backbones, and other modalities. More broadly, it offers a way to characterize training objectives through the geometry of the solutions they reach, rather than through error rates alone.

\section*{AI-Generated Content Disclosure}
The authors used Anthropic's Claude and OpenAI's ChatGPT as writing and editing assistants in preparing this manuscript. The tools were used to draft and revise prose in the abstract, introduction, related work, results, and discussion sections, and to generate and refine the plotting code for the figures from numerical results produced by the authors. All research design, experiments, data, quantitative results, and scientific claims are the authors' own; all AI-assisted text and figures were reviewed, verified, and edited by the authors, who take full responsibility for the content.

\bibliographystyle{IEEEtran}
\bibliography{IEEEabrv,IEEEbiblio}

\end{document}